# Design and Implementation of an Unmanned Vehicle Using A GSM Network without Microcontrollers

**Sourangsu Banerji,**
Department of Electronics & Communication Engineering,
RCC-Institute of Information Technology, India

**ABSTRACT** : In the recent past, wireless controlled vehicles had been extensively used in a lot of areas like unmanned rescue missions, military usage for unmanned combat and many others. But the major disadvantage of these wireless unmanned robots is that they typically make use of RF circuits for maneuver and control. Essentially RF circuits suffer from a lot of drawbacks such as limited frequency range i.e. working range, and limited control. To overcome such problems associated with RF control, few papers have been written, describing methods which make use of the GSM network and the DTMF function of a cell phone to control the robotic vehicle. This paper although uses the same principle technology of the GSM network and the DTMF based mobile phone but it essentially shows the construction of a circuit using only 4 bits of wireless data communication to control the motion of the vehicle without the use of any microcontroller. This improvement results in considerable reduction of circuit complexity and of manpower for software development as the circuit built using this system does not require any form of programming. Moreover, practical results obtained showed an appreciable degree of accuracy of the system and friendliness without the use of any microcontroller.

## 1. INTRODUCTION

In principle, RF (Radio Frequency) can be regarded as the control which deals with the use of radio signals to remotely control any device. A remotely controlled vehicle may be defined as any mobile device which is controlled by means that it does not restrict its motion with an origin external to the device i.e. the possibility of an existence of a radio control device, a cable between the control and the vehicle or an infrared controller. A RCV is always controlled by a human operator and takes no positive action autonomously.

The IR system follows the line of sight approach which involves the process of actually pointing the remote at the device being controlled; this makes communication over obstacles and barrier quite impossible. To overcome such problems, a signaling scheme utilizing voice frequency tones is employed. This is known as Dual Tone Multi-Frequency (DTMF), Touch-Tone or simply tone dialing. A valid DTMF signal is the sum of two tones, one from a low group (697-941Hz) and the other from a high group (1209-1633Hz) with each group containing four individual tones. DTMF signaling therefore play an important role in distributed communication systems such as multiuser mobile radio.

In this paper, phones making use of the GSM network interfaced directly with the DTMF decoder and the motor driver is used to remotely control an unmanned robotic vehicle thus overcoming the distance barrier problem and communication over obstacles with very minimal or no interference but is solely network dependant. The design of an unmanned vehicle proposed here does not make use of any microcontroller. The transmitter used, is a handheld cell phone.

A literature review is given in section 2 followed by our proposed model in section 3. The hardware design framework is discussed in section 4. The circuit design; construction and working are explained in sections 5 and 6. Experimental results, comparison and cost analysis are covered in sections 7, 8 and 9. The applications and future scope is discussed in the sections 10 and 11. Lastly we conclude our paper.

## 2. LITERATURE REVIEW

In 1898, Nikola Tesla built the first propeller-driven radio controlled boat, which can be regarded as the original prototype of all modern-day uninhabited aerial vehicles and precision guided weapons. Records state that it is the first among all remotely controlled vehicles in air, land or sea. It was powered by lead-acid batteries and an electric drive motor. The vessel was designed in such a way that it could be maneuvered alongside a target using instructions received from a wireless remote-control transmitter. Once in its position, a command would be sent to detonate an explosive charge contained within the boat's forward compartment. The weapon's guidance system introduced a secure communications link between the pilot's controller and the surface-running torpedo in an effort to assure that control could be maintained even in the presence of electronic counter measures.

Wireless controlled unmanned vehicles which are used nowadays typically use RF circuits for motion and control. But RF circuits suffer from the disadvantage of limited working range which results in limited control. As RF circuits' follows LOS (Line of sight) approach, it fails miserably in NLOS (Non-Line of Sight) conditions involving obstacles and barriers. To overcome these, one method was proposed by Awab Farikh et al, (2010) [1] which typically makes use of the DTMF technology along with a microcontroller based circuit for maneuver and control of these unmanned robotic vehicles.

Similarly, Ashish Yadav et al, (2012) [2] also proposed the construction of an unmanned vehicle which could be especially used for ground combat using a similar technology. Recently, Sabuj Das Gupta et al (2013) [3] discussed in detail about how the method described in [1] could be implemented using a microcontroller by providing the necessary circuit details and the software code.

We implemented the design of an unmanned vehicle using the same technology as described in the papers mentioned earlier, modified the existing circuit and code described in paper [3] and gave a thorough and detailed analysis of the design paradigm which is the best possible explanation to our knowledge till date in [4].

However, upon implementation we also found that the proposed method could also be implemented without using a microcontroller, which was one of the key elements in the design of the circuit. In case of motion control as described in these papers [1-4], we found that a considerable amount of circuit complexity can be reduced when we omit the use of the microcontroller. And the need for writing any software code is also absent.

## 3. PROPOSED MODEL

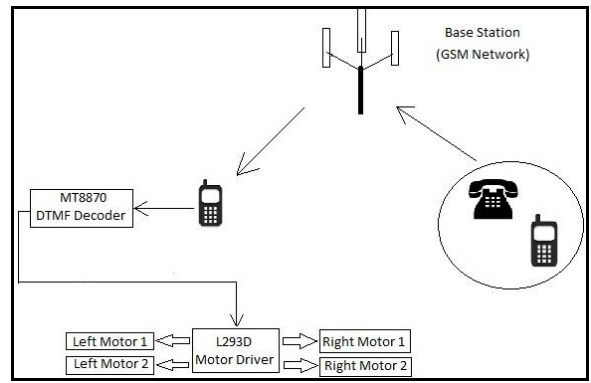

**Figure 1** Functional Block Diagram

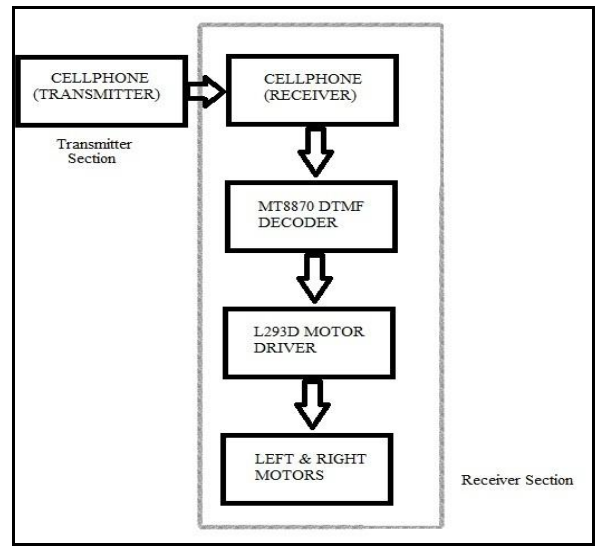

**Figure 2** Block Diagram of the proposed model

The diagrams in fig.1 and fig. 2 describe the overall system. The robotic vehicle is basically operated with the help of any GSM enabled mobile phone which makes a call to the phone stacked on the robot. The phone at the receiver end perceives the DTMF tone from the mobile/cordless phone at the transmitter end and feeds the signal as input to the DTMF decoder. The DTMF decoder processes the tone and feeds the output as input to the motor driver. Accordingly the robotic vehicle operates. The mobile that makes a call to the mobile phone stacked in the vehicle basically acts as a remote. So this simple robotic car does not require the construction of receiver and transmitter units, thereby further reducing the circuit complexity.

## 4. HARDWARE DESIGN FRAMEWORK

The blocks of the receiver model which is seen in fig.2 are explained in detail in this section:

### 4.1 DTMF DECODER

An MT8870 (fig. 3) series DTMF decoder is used here. The MT8870 series use digital counting techniques to detect and decode all the 16 DTMF tone pairs into a 4-bit code output. The need for pre-filtering is eliminated using the built-in dial tone rejection circuit. In single-ended input configuration when the input signal given at pin 2, is effective, the correct 4-bit decoded signal of the DTMF tone is generated and transferred to Q1 (pin 11) through Q4 (pin 14) outputs of the DTMF decoder which are given as input to the corresponding input pins of the motor driver.

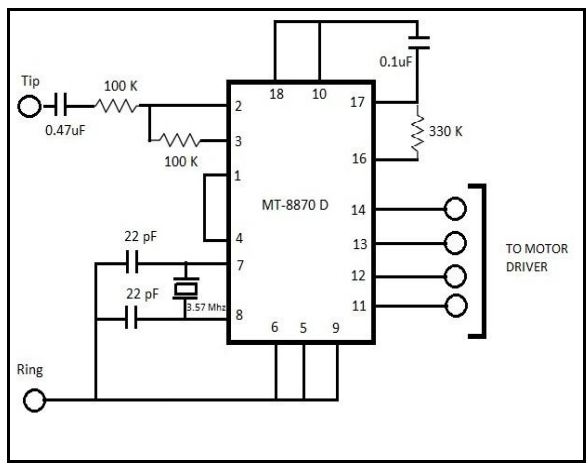

**Figure 3** DTMF decoder configuration

Integration of both the band split filter and digital decoder functions makes the DTMF decoder a complete receiver. Switch capacitor techniques for high and low group filters are used; digital counting techniques are used to detect and decode all 16 DTMF tone pairs into a 4 bit code. The external component count is minimized as an on chip provision of a differential input amplifier clock oscillator and latched three state bus interfaces are provided. The functional description of the MT8870 is given in the following sections:

### 4.1.1 Filter Section:

Separation of the low group and high group tones are achieved by applying the DTMF signal to the input of the two sixth order switched capacitor band pass filter, the band width of whose correspond to the low and high group frequencies. Prior to limiting each filter output is followed by a single order switched capacitor filter section which conditions the signal; limiting is performed by the use of high gain comparators which are provided with hysteresis to prevent detection of unwanted low level signals. Full rail logic swing is provided at the output of the comparators at the frequency of the incoming DTMF signals.

### 4.1.2 Decoder Section:

In the decoder section, there is a decoder employing digital counting techniques to determine the frequencies of the incoming tones and to verify whether they correspond to standard DTMF frequencies or not. Use of a complex averaging algorithm protects against tone simulation by extraneous signals such as voice in addition to providing tolerance for small frequency deviations and variations.

The development of this averaging variation algorithm is to ensure an optimum combination of immunity to talk off and tolerance to the presence of interfering frequencies (third tone) and noise. As the presence of two valid tones is recognized (this is referred to as the signal condition) the early steering ($E_{st}$) output will go in to an active state. Subsequent loss of signal condition, if any, will force $E_{ST}$ to assume an inactive state.

### 4.1.3 Steering Circuit:

The receiver checks for a valid signal duration before registration of a decoded tone pair. An external Resistance Capacitance (RC) of time constant $E_{st}$ performs the check. Logic high on $E_{st}$ causes the collector voltage ($V_c$) to rise as the capacitor discharges. The function of the decode algorithm is to estimate the time required to detect the presence of two valid tones top, the tone frequency and the previous state of the decode logic. $E_{ST}$ indicates and initiates an RC timing circuit.

If both tones are present for the minimum guide time ($t_{CTP}$) which is determined by the external RC network, decoding of the DTMF signal takes place and the resulting data is latched onto the output register. Indication that new data is available is given when the delay steering ($S_{tD}$) output is raised high. The time required to receive a valid DTMF signal ($T_{rec}$) is equal to the sum of time to detect the presence of valid DTMF signals ($t_{DP}$) ,guard time and the tone present.

### 4.2 MOTOR DRIVER

The L293D is a quad, high-current, half-H driver designed to produce two way drive currents of up to 600 mA at voltages ranging from 4.5V to 36V which makes it easier to drive the DC motors (fig.4).

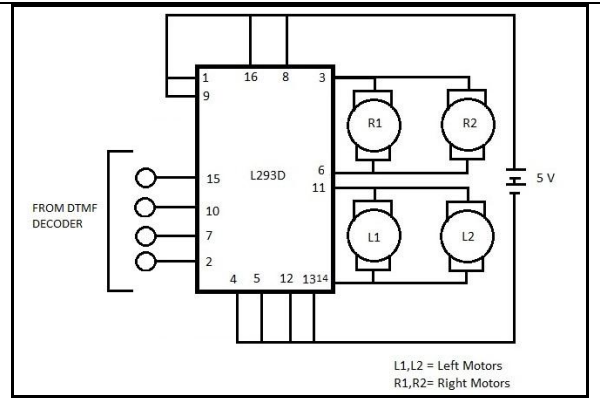

**Figure 4** Motor driver configuration

It basically consists of four drivers. The pins IN1 through IN4 and OUT1 through OUT4 are input and output pins respectively of driver 1 through driver 4. Drivers 1 and 2 as well as drivers 3 and 4 are enabled by enable pin 1 (EN1) and pin 9 (EN2) accordingly. When enable input EN1 (pin 1) is high, drivers 1 and 2 are enabled and the outputs corresponding to their inputs are active high. Similarly, enable input EN2 (pin 9) enables drivers 3 and 4.

## 5. CIRCUIT DIAGRAM

In this section, the circuit diagram of the project is shown below (fig. 5). The construction and the working methodology of the circuit of the unmanned robotic vehicle have been discussed in the following section.

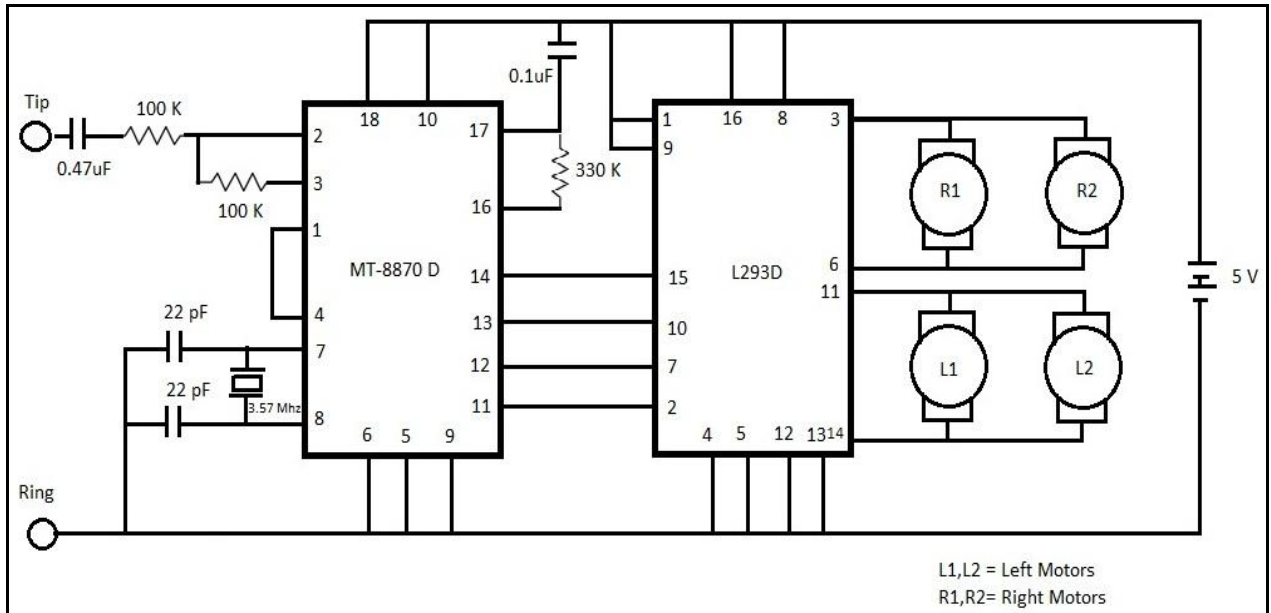

**Figure 5** Circuit Diagram

## 6. CONSTRUCTION & WORKING

The unmanned vehicle could either have a two wheel drive or a four wheel drive. We choose to go with a four wheel drive as it provides a better control and more torque than the two wheel system. The circuit as shown in fig.5 has been designed on a breadboard.

The breadboard on which the circuit was constructed is mounted on a steel chassis. In addition the cell phone (receiver one) which is attached to the vehicle is also mounted on the chassis (not shown in the figure). Motors which are used for motion of the robotic car are fixed to the bottom of the steel chassis. In the four wheel drive system which we had used in the configuration of our circuit, the motors on both the sides are managed independently of one another. However a single L239D motor driver IC is enough to control the four motors. Now to operate the vehicle, we need to make a call to the cell phone attached to the circuit on the receiver side. Moreover, we should note, that the call is only possible if the operator on the transmitter side knows the cell phone number of the other phone. Any GSM enabled phone can be used as the transmitter; which sends the DTMF tones through the existing GSM network.

The tones are received by the receiver phone accordingly. One thing to keep in mind is that the cell phone at the receiver side should be kept in auto answer mode, so that the call can be taken after a single ring. The DTMF tones which are received are fed to the circuit through the headset of the cell phone.

Referring to our circuit diagram (fig.5), the ring and the tip refer to the two wires that go into them, which we will get when we will split open the headset wire of our cell phone. If we cut open the ear phone wire we will see three wires coming out. They are:
1. Red wire (headset right output-Ring),
2. Blue wire (headset left output-Tip) and
3. Copper wire (ground wire).

The wires are laminated and the lamination should be removed before they are inserted into the ring and tip positions as shown in the circuit diagram. If this procedure is not followed, the tones may not be received by the DTMF decoder at all or a very weak signal would be received by the decoder, which cannot be easily decoded by it [4].

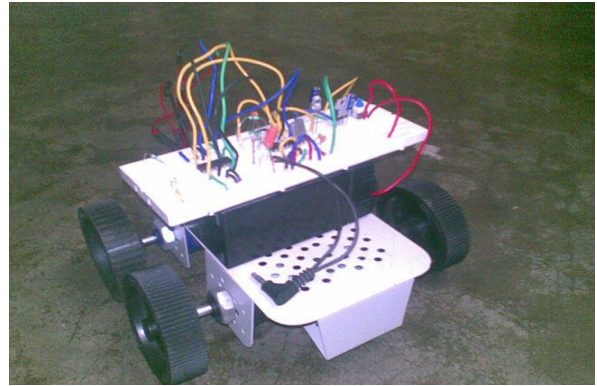

**Figure 6** Working Model

The DTMF tones are fed as input to the DTMF decoder which produces the 4 bit equivalent of the received tone and passes it on to the motor driver as input. The motor driver on receiving the corresponding input drives the motor as specified.

We see that when key '6' is pressed on the mobile phone, the robotic car executes forward motion. When we press key '9' on our mobile phone, the car moves in the reverse direction. When we press key '4', the car takes a left turn. When we press key '2' on our mobile phone, the vehicle turns in the right direction. Similarly when '0' is pressed on the cell phone, the vehicle halts. Five keys on the keypad are used for motion control of the unmanned car.

## 7. EXPERIMENTAL RESULTS

The experimental values of the frequency of sinusoidal waves for DTMF and the voltage level at the out-put pins of MT8870 and L293D motor driver were measured. These values are juxtaposed with the theoretical values in the Table 1 and 2 respectively. Table 3 shows the HEX reading obtained from output pins of MT8870 and L293D. Table 4 shows how the robot's resultant movement is achieved as a combination of the right and left motor movements. We also compared the output value of the motor driver given in Table 5 when input was given from:
1. Microcontroller and
2. DTMF decoder directly.

**TABLE 1**
Frequency Readings

| KEY | LOWER FREQUENCY (Hz) | | HIGHER FREQUENCY(Hz) | |
|---|---|---|---|---|
| | TH. | EXP. | TH. | EXP. |
| 2 | 697 | 672 | 1336 | 1320 |
| 4 | 770 | 731 | 1209 | 1201 |
| 6 | 770 | 731 | 1477 | 1475 |
| 8 | 852 | 855 | 1336 | 1322 |
| 5 | 770 | 735 | 1336 | 1325 |

**TABLE 2**
Voltage Readings

| LOGIC LEVEL | OUPUT VOLTAGE OF MT8870 | | OUTPUT VOLTAGE OF L293D | |
|---|---|---|---|---|
| | TH. | EXP. | TH. | EXP. |
| LOW | 0.03 | 0.09 | 0 | 0.11 |
| HIGH | 4.97 | 4.80 | 5 | 4.82 |

**TABLE 3**
Hex readings corresponding to robot's movement

| KEY PRESSED | OUTPUT OF MT8870 | INPUT OF L293D | ROBOT'S NET MOVEMENT |
|---|---|---|---|
| 6 | 0110 | 0110 | Forward |
| 9 | 1001 | 1001 | Left turn |
| 4 | 0100 | 0100 | Right turn |
| 2 | 0010 | 0010 | Backward |
| 0 | 0000 | 0000 | Stop |

**TABLE 4**
Combination of Left and Right Motor Movement

| INPUT | RIGHT MOTOR ACTION | LEFT MOTOR ACTION | RESULTANT MOTION |
|---|---|---|---|
| 0110 | Forward | Forward | Forward |
| 1001 | Reverse | Reverse | Reverse |
| 0100 | Forward | Stop | Left |
| 0010 | Stop | Forward | Right |
| 0000 | Stop | Stop | Stop |

**TABLE 5**
Output of L293D Motor Driver

| MOTORS | VOLTAGE ACROSS MOTOR | |
|---|---|---|
| | FROM MICRO CONTROLLER | FROM DTMF DECODER |
| LEFT(MOTION) | 4.20 | 4.80 |
| LEFT(REST) | 0.01 | 0.09 |
| RIGHT(MOTION) | 4.20 | 4.80 |
| RIGHT(REST) | 0.01 | 0.09 |

## 8. COST ANALYSIS

Cost analysis is an integral part of any project that is carried out. In this section, we try to provide an approximate cost estimate of the project. The components used here can be brought from any of the online electronics stores which are found on the internet. Although, most of the components used in this project can be found in any college laboratory. A list of the components and their prices are given in the following table (Table 6):

**TABLE 6**
Cost of the Components

| Components | Quantity | Cost |
|---|---|---|
| MT8870 I.C. | 1 | $1.95 |
| L293D I.C. | 1 | $1.50 |
| 100kilo-ohm Resistor | 2 | College Lab |
| 330kilo-ohm Resistor | 1 | College Lab |
| 0.47 uF | 1 | College Lab |
| 22pF | 2 | College Lab |
| 0.1uF | 1 | College Lab |
| Crystal (3.57MHz) | 1 | $0.15 |
| Geared Motors | 4 | $12 |
| Breadboard | 1 | College Lab |
| Battery(6V) | 1 | $4.50 |
| | Total | $20.1 |

From the table above we can see that this unmanned vehicle can be made with components which are cheap and very readily available in the market. The circuit was designed keeping in mind the implementation cost considerations. This total project cost is around or under $20 and if manufactured in bulk the cost will be even less.

## 9. COMPARISION

In the previous two sections, we have seen that our proposed model achieves a considerable amount of accuracy as is evident from Table 5 in case of motion control. More specifically looking at the circuit diagram it is clearly visible that our circuit requires much less components to obtain the same objective as is proposed in papers [2-4]. From convention reduced circuitry results in less complexity.

A cost analysis shows that our proposed design method is also quite cost effective. This reduces manufacturing costs also. There is no need of any programming so deployment of manpower to program the circuit is also not required.

## 10. APPLICATION

These unmanned vehicles can have various scientific uses in hazardous and unknown environments. USVs have also been used for space exploration purposes examples of which are Voyager-I and the Martian explorers Spirit and Opportunity. Similarly, military usage of these robotic vehicles dates back to the first half of the $20^{th}$ century [6].

Remote controlled vehicles are used by many police department bomb-squads to defuse or detonate explosives. Current USV's can hover around possible targets until they are positively identified before releasing their payload of weaponry. USVs also play an increased role in search and rescue. These vehicles could be used in case of natural calamities & emergencies. This can also be a great asset to save the lives of both people and soldiers.

In recent times, there has been a serious endangerment to the wildlife. Many animals are significantly on the verge of extinction. These spy robotic cars can be used to patrol the different sections of the forested areas for any suspicious activity and since it is a live streaming device as well as mobile, it can keep the forest guards constantly updated and alert about the status of different areas which are vulnerable to attack. As a result, it can help to prevent further destruction of the forest resources by enabling correct prohibitory action at the appropriate time.

## 11. FUTURE SCOPE

As the microcontroller which is essentially "a micro-computer on a chip" is not used we are truly unaware and not certain at this point of time whether extra functions can be added to the robotic vehicle in addition to direction control. Considerable amount of intensive research in this area is required to come to any such definite conclusion. Even if it is possible then as described in paper [4] the project can be extended to include IR sensors, a camera or even a system for password protection of the USV. Now, IR sensors can be automatically used to detect and avoid obstacles if the vehicle goes beyond the line of sight. Coming to password protection, a brilliant scheme has already been described in paper [5] where the purpose had been achieved without the use of any microcontroller. Moreover, still research is needed to see whether the inclusion of a camera with the vehicle could be achieved without the use of a microcontroller and without much circuit complexity.

## 12. CONCLUSION

The basic aim to develop an USV which overcomes the drawbacks of the conventionally used RF circuits has already been achieved in papers [1-2]. But the method proposed in this paper can be said to improve upon the current design or provide an alternative to the existing one. The design described here does not make use of any microcontroller and it can be seen that even without the usage of the microcontroller we achieve similar and comparable accuracy in direction control. Whether new functions which are described in the section 11 can be added is beyond the scope of this paper. As discussed in section 9, the advantage of this system can be summarized as reduced circuit complexity, reduced cost of manufacturing and ease of deployment. In addition to these advantages, the robotic car constructed based on the design in this paper would have the same benefits associated with the designs described with microcontroller.

*Acknowledgements*

We would like to thank Prof. Saraswati Saha, Department. Of Electronics & Communication Engineering, RCC-Institute of Information

## References

1. Awab Fakih, Jovita Serrao, "Cell Phone Operated Robotic Car."International Journal of Scientific & Engineering Research, ISSN 2229-5518.
2. Gupta, Sabuj Das, Arman Riaz Ochi, Mohammad Sakib Hossain, and Nahid Alam Siddique. "Designing & Implementation of Mobile Operated Toy Car by DTMF."International Journal of Scientific& Research Publications, Vol.-3, Issue-1,2013.ISSN 2250-3153.
3. Jadhav, Ashish, Mahesh Kumbhar, and Meenakshi Pawar. "Cell Phone Controlled Ground Combat Vehicle." International Journal of Computer and Communication Engineering, Vol. 1, No. 2, July 2012.
4. Banerji, Sourangsu. "Design and Implementation of an Unmanned Vehicle using a GSM network with Microcontrollers." International Journal of Science, Engineering and Technology Research 2.2 (2013): pp-367-374.
5. Shashanka, D. "Password Protection for DTMF Controlled Systems without Using a Microcontroller."International Conference on Computing & Control Engineering,2012.
6. Edwin Wise, Robotics Demystified (Mc-Graw Hill, 2005)